\documentstyle[psfig]{mn2e}

\def\ga{\mathrel{\hbox{\rlap{\hbox{\lower4pt\hbox{$\sim$}}}\hbox{$>$}}}}
\def\fd{\hbox{$.\!\!^{\rm d}$}}

\title[High-speed photometry of faint Cataclysmic Variables: IV.]
{High speed photometry of faint Cataclysmic Variables: IV. 
V356 Aql, Aqr1, FIRST\,J1023+0038, H$\alpha$\,0242--2802, GI Mon, AO Oct, V972 Oph, 
SDSS\,0155+00, SDSS\,0233+00, SDSS\,1240-01, SDSS\,1556-00, SDSS\,2050-05, FH Ser}
\author[Patrick A. Woudt, Brian Warner and Magaretha L. Pretorius]
       {Patrick A. Woudt\thanks{E-mail: pwoudt@circinus.ast.uct.ac.za}, 
        Brian Warner\thanks{E-mail: warner@physci.uct.ac.za},
        Magaretha L. Pretorius\thanks{E-mail: retha@mensa.ast.uct.ac.za}\\
        Department of Astronomy, University of Cape Town, Private Bag,
        Rondebosch 7700, South Africa}
\date{}

\begin{document}

\maketitle

\begin{abstract}
   We present results from high speed photometry of a further thirteen faint 
cataclysmic variables. V356 Aql (Nova Aql 1936) shows flare-like outbursts 
with recurrence time scales $\sim$ 3000 s, but no coherent periodicities. Aqr1 is 
an intermediate polar with a spin period of 6.7284 min and a probable orbital period 
$P_{orb}$ = 3.226 h derived from orbital sidebands. Its orbital sideband frequencies show very variable 
amplitudes. The published spectroscopic period of 2.0 h suggests that Aqr1 is similar to GW Lib and FS Aur
in having an additional periodicity of unknown origin. FIRST\,J1023+0038 has $P_{orb}$ = 4.7548 h with an orbital 
modulation of range 0.45 mag, probably caused by reflection effect from a 
hot white dwarf primary; as such it may have been a nova sometime in the 
past few decades. H$\alpha$\,0242 is a deeply eclipsing very low mass-transfer rate 
system, probably a dwarf nova of very long outburst interval, with $P_{orb}$ = 
1.792 h. GI Mon, an old nova, has optical modulations at 4.325 h and possibly also at 48.6 min and is thus
a candidate intermediate polar. AO Oct, an SU UMa type dwarf nova, shows orbital modulation at 
quiescence with $P_{orb}$ = 94.097 min. V972 Oph (Nova Ophiuchi 1957) showed 
no flickering activity during one set of observations, but did so at a later 
time, confirming the correctness of the identification of this object, but it 
shows no orbital modulation. SDSS\,0155 is a deeply eclipsing polar with an orbital
period of 87.13 min. SDSS\,0233 shows flaring activity but no discernible periodicity.
No periodicity was found in the new AM CVn candidate SDSS\,1240.
SDSS\,1556 shows a periodic modulation at 
1.778 h which is possibly due to orbital motion. SDSS\,2050 is an eclipsing 
polar with $P_{orb}$ = 1.5702 h. 
FH Ser (Nova Serpentis 1970) has strong flickering activity but no detectable 
orbital modulation. 
\end{abstract}


\section{Introduction}

    Continuing the University of Cape Town survey of faint cataclysmic 
variable stars (CVs) that have not previously been studied with high time 
resolution photometry, we present results for a further 13 stars. The 
observational technique is the same as that employed previously: the UCT 
CCD photometer (O'Donoghue 1995) used in frame transfer mode and 
white light, attached to the 1.9-m (74-in) or 1.0-m (40-in) reflectors at the 
Sutherland site of the South African Astronomical Observatory. In addition a 
few of our runs were made with the SAAO CCD photometer on the 1.0-m 
reflector. With white light, and with the very non-standard flux distributions 
of CVs, no transformations onto recognised photometric systems are 
possible -- but in any case we are largely interested in the brightness 
variations that occur on short time scales, and which are measured 
differentially with respect to the mean brightness of each variable star. 
Nevertheless, by observing hot white dwarf standards we are able to provide 
magnitudes that are roughly on the V scale, and are good to $\sim$ 0.1 mag.

    In the previous papers (Woudt \& Warner 2001, 2002a, 2003a) we have 
concentrated on faint nova remnants. Here we also give results for some old 
novae, but we have extended our coverage to include a number of faint CVs 
that have been found in recent large-scale surveys.

    In Section 2 we give the detailed results of our observations. Table 1 
contains the observing log for all of the runs on the various observed stars. 
Section 3 summarises the results.

\section{Observations}

\begin{table*}
 \centering
  \caption{Observing log.}
  \begin{tabular}{@{}llrrrrrcc@{}}
 Object       & Type         & Run No.  & Date of obs.          & HJD of first obs. & Length    & $t_{in}$ & Tel. &  V \\
              &              &          & (start of night)      &  (+2452000.0)     & (h)       &     (s)   &      & (mag) \\[10pt]
{\bf V356 Aql}& NR           & S6251    & 2001 Sep 24 & 2177.23744  &    2.43     &  20, 40   &  40-in & 17.7 \\
              &              & S6254    & 2001 Oct 09 & 2192.23163  &    3.56     &      20   &  40-in & 17.8 \\
              &              & S6258    & 2001 Oct 10 & 2193.22754  &    2.80     &  20, 40   &  40-in & 17.8 \\
              &              & S6262    & 2001 Oct 11 & 2194.22808  &    3.34     &  30, 40   &  40-in & 17.9 \\
              &              & S6266    & 2001 Oct 12 & 2195.22790  &    2.32     &  20, 40   &  40-in & 17.8 \\
              &              & S6279    & 2001 Oct 19 & 2202.23856  &    2.20     &       8   &  74-in & 17.8 \\
              &              & S7061    & 2003 Aug 28 & 2880.21490  &    2.40     &      45   &  40-in & 17.8$^b$ \\[5pt]
{\bf Aqr1}    & IP           & S6600    & 2002 Oct 31 & 2579.32931  &    2.30     &      20   &  74-in & 18.4 \\
              &              & S6602    & 2002 Nov 01 & 2580.25848  &    4.98     &      20   &  74-in & 18.3 \\
              &              & S6604    & 2002 Nov 02 & 2581.25284  &    4.98     &      20   &  74-in & 18.3 \\
              &              & S6608    & 2002 Nov 04 & 2583.25411  &    4.76     &      20   &  74-in & 18.2 \\[5pt]
{\bf First}   & P?           & S6708    & 2002 Dec 30 & 2639.47195  &    3.30     &  10, 15   &  74-in & 17.6 \\
              &              & S6711    & 2003 Jan 28 & 2668.39976  &    5.01     &      20   &  40-in & 17.5 \\
              &              & S6714    & 2003 Jan 29 & 2669.45782  &    3.90     &      20   &  40-in & 17.5 \\
              &              & S6716    & 2003 Jan 30 & 2670.40913  &    4.91     &      20   &  40-in & 17.5 \\
              &              & S6719    & 2003 Jan 31 & 2671.38651  &    4.70     &      20   &  40-in & 17.5 \\
              &              & S6736    & 2003 Feb 03 & 2674.56766  &    1.26     &      20   &  40-in & 17.6 \\
              &              & S6744    & 2003 Feb 05 & 2676.52060  &    2.76     &      25   &  40-in & 17.6 \\
              &              & S6785    & 2003 Feb 21 & 2692.30455  &    4.08     &      25   &  40-in & 17.5 \\
              &              & S6803    & 2003 Feb 24 & 2695.45347  &    3.23     &      25   &  40-in & 17.6 \\[5pt]
{\bf H{$\alpha$}\,0242--2802}& DN & S6591  & 2002 Oct 29 & 2577.37719  &    3.69     &  45, 90   &  74-in & 18.7:$^*$\\
              &              & S6595    & 2002 Oct 30 & 2578.46831  &    0.56     &      25   &  74-in & 18.7:$^*$\\
              &              & S6597    & 2002 Oct 30 & 2578.53360  &    0.95     &      45   &  74-in & 18.7:$^*$\\[5pt]
{\bf GI Mon}  & NR           & S6142    & 2000 Dec 21 & 1900.45337  &    1.21     &       5   &  40-in & 16.3 \\
              &              & S6155    & 2000 Dec 25 & 1904.36811  &    5.43     &      10   &  40-in & 16.4 \\
              &              & S7227    & 2003 Dec 30 & 3004.40587  &    4.60     &      30   &  40-in & 16.5$^a$ \\
              &              & S7228    & 2003 Dec 31 & 3005.35415  &    4.91     &      30   &  40-in & 16.5$^a$ \\
              &              & S7230    & 2004 Jan 01 & 3006.35296  &    5.98     &      30   &  40-in & 16.4$^a$ \\[5pt]
{\bf AO Oct}  & DN           & S6102    & 2000 Jun 05 & 1701.51362  &    4.02     & 60, 120   &  74-in & 20.5 \\
              &              & S6121    & 2000 Aug 22 & 1779.39636  &    6.62     &45, 60, 120&  74-in & 20.1 \\
              &              & S6125    & 2000 Aug 24 & 1781.50967  &    3.90     &  45, 60   &  74-in & 20.2 \\
              &              & S6131    & 2000 Aug 27 & 1784.47835  &    3.85     &      60   &  74-in & 20.0 \\[5pt]
{\bf V972 Oph}& NR           & S6342    & 2002 Apr 02 & 2367.45468  &    1.29     &   6, 15   &  74-in & 15.9 \\
              &              & S7064    & 2003 Aug 29 & 2881.21042  &    4.66     &      30   &  40-in & 16.1$^a$ \\
              &              & S7071    & 2003 Aug 30 & 2882.21084  &    2.56     &      30   &  40-in & 16.2$^a$ \\[5pt]
{\bf SDSS\,0155+00}& P       & S7141    & 2003 Oct 04 & 2917.39123  &    0.85     &      45   &  74-in & 18.0$^*$ \\
              &              & S7143    & 2003 Oct 05 & 2918.36033  &    1.76     &  10, 45   &  74-in & 18.0$^*$ \\[5pt]
{\bf SDSS\,0233+00}& DN?     & S6519    & 2002 Aug 31 & 2518.52361  &    1.65     &      90   &  74-in & 19.9 \\
              &              & S6525    & 2002 Sep 01 & 2519.48797  &    1.65     &      90   &  74-in & 19.9 \\
              &              & S6527    & 2002 Sep 01 & 2519.64325  &    0.60     &     120   &  74-in & 19.8 \\
              &              & S6594    & 2002 Oct 30 & 2578.30153  &    3.83     &     100   &  74-in & 19.8 \\
              &              & S7116    & 2003 Sep 23 & 2906.54136  &    2.10     &      60   &  74-in & 19.9 \\
              &              & S7119    & 2003 Sep 24 & 2907.50635  &    2.03     &     100   &  74-in & 19.9 \\[5pt]
{\bf SDSS\,1240-01}& AM CVn  & S7240    & 2004 Feb 14 & 3050.50814  &    1.60     &      90   &  40-in & 19.6 \\
              &              & S7246    & 2004 Feb 16 & 3052.45836  &    0.83     &     120   &  40-in & 19.6 \\
              &              & S7254    & 2004 Feb 19 & 3055.50047  &    2.13     &      90   &  74-in & 19.6 \\
              &              & S7257    & 2004 Feb 20 & 3056.50522  &    1.22     &      90   &  74-in & 19.6 \\
              &              & S7260    & 2004 Feb 21 & 3057.55213  &    2.20     &     100   &  74-in & 19.7 \\[5pt]
{\bf SDSS\,1556-00}& DN      & S6812    & 2003 Feb 26 & 2697.51362  &    0.70     &      60   &  40-in & 17.9 \\
              &              & S6870    & 2003 Mar 26 & 2725.49996  &    3.08     &      45   &  74-in & 18.0 \\
              &              & S6872    & 2003 Mar 27 & 2726.48824  &    2.09     &      45   &  74-in & 18.0 \\
              &              & S6881    & 2003 Mar 29 & 2728.46586  &    2.06     &      45   &  74-in & 18.0 \\
              &              & S6887    & 2003 Mar 30 & 2729.43842  &    2.16     &      45   &  74-in & 18.0 \\[5pt]
\end{tabular}
{\footnotesize 
\newline 
Notes: NR = Nova Remnant; IP = Intermediate Polar; P = Polar; DN = Dwarf Nova; $t_{in}$ is the integration time; `:' denotes an uncertain 
value; $^*$ mean magnitude out of eclipse; $^a$ taken with the SAAO CCD; $^b$ taken with a V filter.\hfill}
\label{tab1}
\end{table*}
\addtocounter{table}{-1}

\begin{table*}
 \centering
  \caption{Continued: observing log.}
  \begin{tabular}{@{}llrrrrrcc@{}}
 Object       & Type         & Run No.  & Date of obs.          & HJD of first obs. & Length    & $t_{in}$ & Tel. &  V \\
              &              &          & (start of night)      &  (+2452000.0)     & (h)       &     (s)   &      & (mag) \\[10pt]
{\bf SDSS\,2050-05}& P       & S7062    & 2003 Aug 28 & 2880.32638  &    3.31     &      30   &  40-in & 18.1$^{*,a}$ \\
              &              & S7065    & 2003 Aug 29 & 2881.41126  &    0.88     &      30   &  40-in & 18.2$^{*,a}$ \\
              &              & S7080    & 2003 Sep 03 & 2886.31274  &    1.95     &      30   &  40-in & 17.7$^*$ \\
              &              & S7091    & 2003 Sep 16 & 2899.23195  &    1.13     &      60   &  40-in & 17.8$^*$ \\
              &              & S7093    & 2003 Sep 17 & 2900.26348  &    0.62     &      60   &  40-in & 17.8$^*$ \\
              &              & S7099    & 2003 Sep 20 & 2903.22917  &    1.60     &      60   &  40-in & 18.0$^*$ \\
              &              & S7105    & 2003 Sep 21 & 2904.31274  &    1.05     &      60   &  40-in & 18.0$^*$ \\
              &              & S7128    & 2003 Sep 29 & 2912.23631  &    0.73     &      30   &  74-in & 18.0$^*$ \\[5pt]
{\bf FH Ser}  & NR           & S7045    & 2003 Aug 22 & 2874.23498  &    3.49     &      10   &  40-in & 17.2:\\
              &              & S7050    & 2003 Aug 23 & 2875.21135  &    3.85     &      10   &  40-in & 17.0:\\
              &              & S7056    & 2003 Aug 27 & 2879.23047  &    4.46     &      30   &  40-in & 17.0$^{a,b}$ \\[5pt]
\end{tabular}
{\footnotesize 
\newline 
Notes: NR = Nova Remnant; P = Polar; $t_{in}$ is the integration time; `:' denotes an uncertain 
value; $^*$ mean magnitude out of eclipse; $^a$ taken with the SAAO CCD; $^b$ taken with a V filter.\hfill}
\label{tab1b}
\end{table*}

\subsection{V356 Aquilae}

V356 Aql was Nova Aquilae 1936, No. 1, discovered by Tamm (1936), 
reaching maximum brightness at $m_{pg}$ = 7.7 and receiving classification as a 
slow nova. It settled down at a post-eruption brightness of V = 18.3 (Szkody 
1994). It has no determination of orbital period, and has not been studied 
with time resolved photometry. The spectrum in the quiescent state 
(Ringwald, Naylor \& Mukai~1996) shows weak H$\alpha$ emission on a continuum that rises 
rapidly at short wavelengths – probably indicative of a high mass transfer 
accretion disc of modest inclination. With the measured H$\alpha$ equivalent width 
of 15 {\AA} and the assumption of an accretion disc background luminosity, an 
inclination $\sim 50^\circ$ is deduced (see Figure 2.20 of Warner 1995).

\begin{figure}
\centerline{\hbox{\psfig{figure=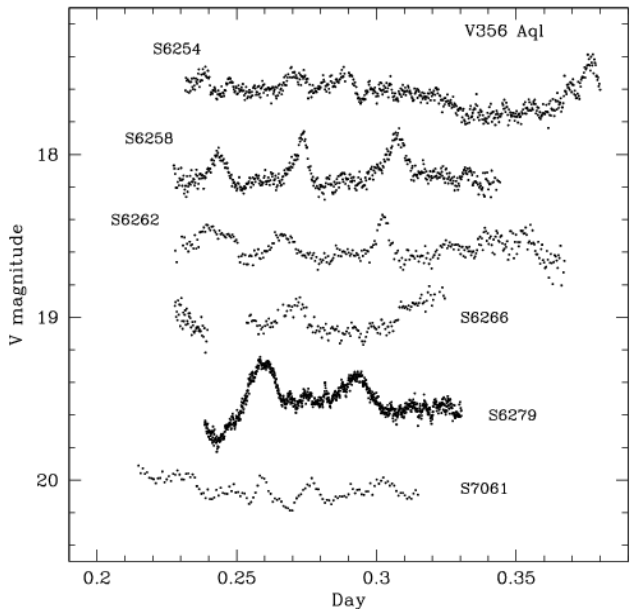,width=8.3cm}}}
  \caption{The light curves of V356 Aql. Runs S6258, S6262, S6266, S6279 and S7061 have been
displaced vertically downwards by 0.35, 0.7, 1.2, 1.7 and 2.3 mag, respectively, for display purposes only.}
 \label{lcv356aql}
\end{figure}

   Our observational runs on V356 Aql are listed in Table 1; we observed it 
on 6 nights in September/October 2001 and then revisited in August 2003 to 
check whether it showed the same behaviour. The light curves are given in 
Fig.~\ref{lcv356aql}. We have been unable to interpret these in any useful way. 
The light curves of V356 Aql during runs S6258, S6262 and S6279 (see Fig.~\ref{lcv356aql})
show flare-like outbursts of $\sim$ 0.3 mag 
recurring on time scales $\sim$ 3000 s, but there is no definite periodicity, and the 
other nights show different behaviours. The FTs quite naturally show excess 
power at $\sim$ 3000 s and harmonics, but there is no coherent period. There 
are no periodicities at shorter time scales.

   There are some similarities of the light curves to those of large amplitude 
ZZ Cet stars (e.g. Warner \& Nather 1972), but the time scale is much longer 
and the surface temperature of a white dwarf that erupted only 70 years ago 
should be much hotter than the ZZ Cet instability strip. Because of accretion 
of hydrogen, the hotter helium-rich pulsational instability strip should not be 
relevant to a nova remnant like V356 Aql.

   Time-resolved spectroscopy used to investigate the changes that take place 
during the flare-like outbursts could help to interpret this unusual object.

\subsection{Aqr1 (= SDSS\,J223843.84+010820.7)}

Listed in the Downes et al.~(2001) catalogue as Aqr1, this star was identified 
as a CV by Berg et al.~(1992) in a survey for bright QSOs, where it was 
designated 2236+0052. It was rediscovered in the Sloan Digital Sky Survey 
(Szkody et al.~2003), where it is listed as SDSS\,2238. The spectrum has 
strong Balmer emission on a blue continuum and shows HeII of moderate 
strength.

   Szkody et al.~(2003) obtained 7 spectra in a 1.7 h run and deduced $P_{orb}$ = 
2.0 h but this must be somewhat uncertain because of the undersampling; they 
also obtained 3.75 h of photometry, with 10 min integrations, and described 
the result as random variability at the 0.02 mag level.

\begin{figure}
\centerline{\hbox{\psfig{figure=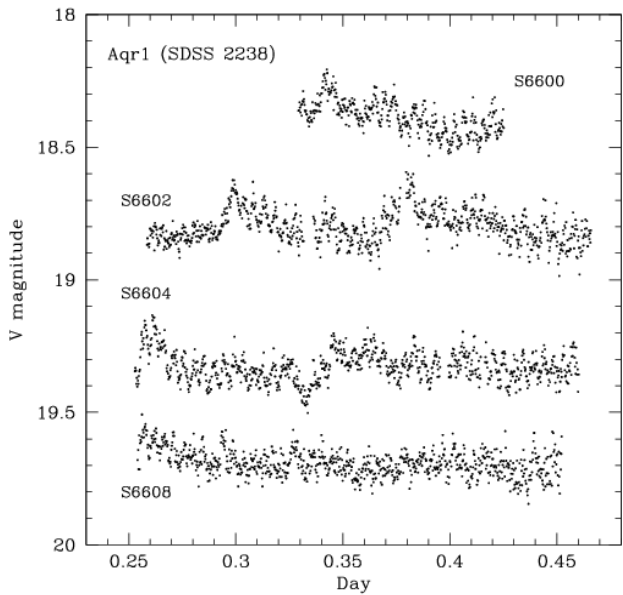,width=8.3cm}}}
  \caption{The light curves of Aqr1 (SDSS\,2238). Runs S6602, S6604 and S6608 have been
displaced vertically downwards by 0.5, 1.0 and 1.5 mag, respectively.}
 \label{lcaqr1}
\end{figure}

\begin{figure}
\centerline{\hbox{\psfig{figure=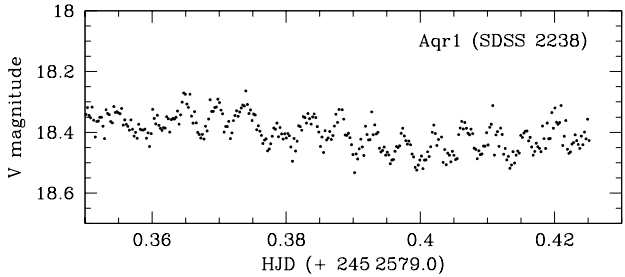,width=8.3cm}}}
  \caption{A detailed view of part of the light curve of run S6600 of Aqr1 (SDSS\,2238).}
 \label{lcaqr1bu}
\end{figure}

   Our observations of Aqr1 are listed in Table 1 and the light curves shown 
in Figs.~\ref{lcaqr1} and \ref{lcaqr1bu}. The light curves show that Aqr1 has a brightness modulation with a 
time scale of several hours and variable range up to 0.2 mag, upon which is 
superimposed a regular periodicity near 6.7 min and maximum range $\sim$ 0.10 
mag. Fig.~\ref{lcaqr1bu} shows a blow up of part of the light curve of run S6600 --
the 6.7-min periodicity is clearly visible in the light curve. This modulation 
is the signature of an intermediate polar (IP). However, there 
is some ambiguity in the interpretation of this star, so we give more details 
of the analysis.

\begin{figure}
\centerline{\hbox{\psfig{figure=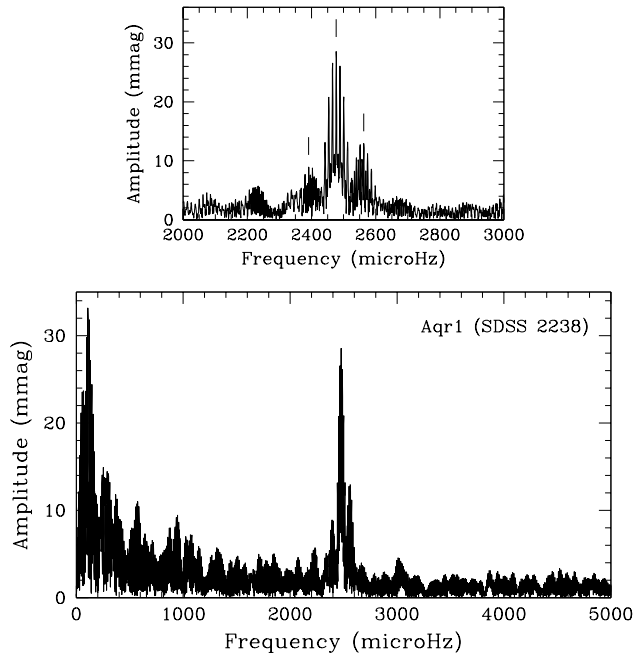,width=8.3cm}}}
  \caption{The Fourier transform of the four long runs on Aqr1. The top panel shows a detailed
section of the Fourier transform around the high frequency modulation.}
 \label{ftaqr1}
\end{figure}

The shorter period appears not to be significantly aliased in the FT of the combined four
runs (upper panel of Fig.~\ref{ftaqr1}), and is at $P$ = 6.72841($\pm$ 16) min with a mean amplitude 0.028 mag. 
There are evident sidebands to the main peak, both at lower and higher 
frequencies. Keeping in mind that in other IPs the sideband frequencies are 
displaced from the main peak by multiples of the orbital frequency $\Omega$ (e.g. 
Warner 1986), we looked for such a suite of frequencies in Aqr1 and met 
with an unexpected problem: in the total FT there is no set of frequencies 
that produces a convincing fit.
There is a strong peak at low frequency (lower panel of Fig.~\ref{ftaqr1}) corresponding
to 2.295 h, with approximately one-day aliases at 2.08 h and 2.55 h. These are near to the spectroscopic
period, but are primarily due to the two strong flares seen in run S6602 (Fig.~\ref{lcaqr1}). If we exclude run
S6602 from the FT, the peak at 2.295 h disappears. In the FT with run S6602 excluded, a low frequency peak
with a fair number of possible aliases is seen around 86 $\mu$Hz, matching the orbital sidebands noted previously.

\begin{figure}
\centerline{\hbox{\psfig{figure=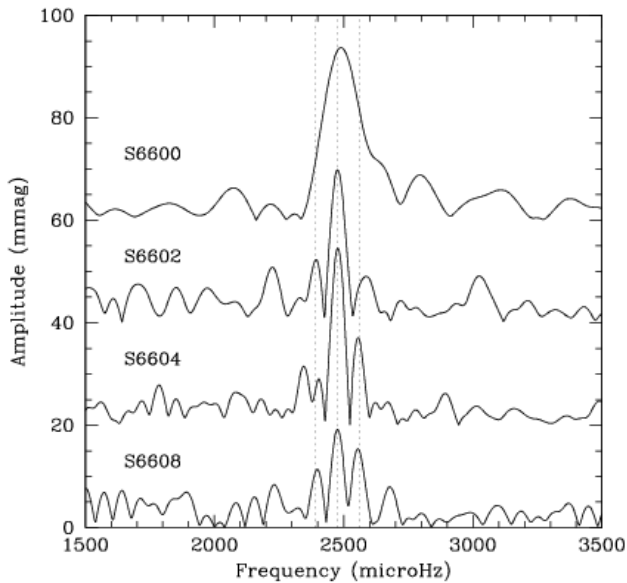,width=8.3cm}}}
  \caption{The Fourier transforms of the four individual runs on Aqr1 (SDSS\,2238), centred on the high frequency
modulation. The Fourier transform of runs S6600, S6602 and S6604 have been displaced vertically by 60, 40 and 20 mmag,
respectively, for display purposes.}
 \label{ftaqr1ind}
\end{figure}

   Denoting the rotation frequency of the primary as $\omega$, these sidebands 
qualitatively fit the scheme (Warner 1986) $\omega - \Omega$, $\omega$, $\omega + \Omega$. There is some indication
in the FT of run S6604 (Fig.~\ref{ftaqr1ind}) for a sideband at $\omega - 2 \Omega$, which suggests
the principal peak as the rotational frequency $\omega$ (in some IPs the 
strongest modulation in the optical is at $\omega - \Omega$). 

We find from a non-linear least squares fit of sinusoids 
that $\omega - \Omega$ = 2391.05 $\mu$Hz, and $\omega + \Omega$ = 2563.19 $\mu$Hz. The mean value of $\Omega$ from the orbital
sidebands is 86.1 $\mu$Hz. These peaks are marked by vertical bars in the upper 
panel of Fig.~\ref{ftaqr1}. Although far less probable, the one-day alias of the orbital frequency at 
74.5 $\mu$Hz cannot be excluded from these data.

Furthermore, we noticed that adding the fourth 
night of data to the first three produced confusion (in the sense of extra 
structure within the sideband envelopes) in the FT. This led us to suspect 
that the sideband amplitudes vary from night to night, which the FT will 
interpret as amplitude modulation and generate appropriate extra sidebands, 
typically with 1/(2 d) spacings, which are what we see within the sideband 
patterns.

    That this is the case is seen in Fig.~\ref{ftaqr1ind}, where we show FTs for the 
individual nights. Although the frequency resolution is low the relative 
amplitudes of the sidebands vary greatly: in S6600 the high frequency 
sideband is not resolved but has about one third of the amplitude of the main 
peak and the low frequency sideband is not clearly present; in S6602 both 
the high and the low frequency sidebands are of low amplitude; in S6604 the 
high frequency sideband is at about half the amplitude of the main peak, and 
there are two bands on the low frequency side of which the furthest is not 
seen in the other runs and appears at about twice the spacing of the usual 
sideband; in S6608 the high frequency sideband is at about 80\% of the 
height of the main peak and the low frequency sideband is about 55\% of the 
main peak.

\begin{figure*}
\centerline{\hbox{\psfig{figure=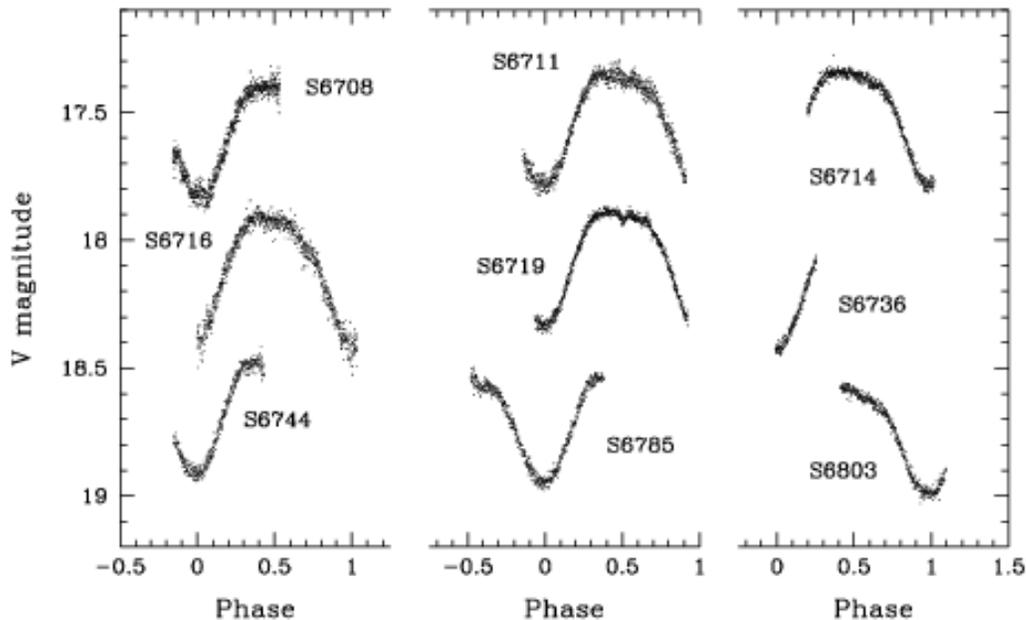,width=16.0cm}}}
  \caption{The light curves of FIRST\,J1023+0038, obtained in 2003 January and February, plotted according to the ephemeris
given in Eq.~\ref{ephfirst}. Light curves for runs S6708, S6711 and S6714 are at the correct brightness, light curves for runs
S6716, S6719 and S6736 have been displaced vertically downwards by 0.6 mag, and light curves for runs S6744, S6785 and S6803 have
been shifted downwards by 1.2 mag for display purposes.}
 \label{lcfirst}
\end{figure*}

We are now confronted by a contradiction: in standard IPs the sideband frequency
splitting agrees exactly with the orbital frequency ($\Omega$), but in Aqr1 our deduced
orbital period from the 86.1 $\mu$Hz splitting is $P_{orb}$ = 3.226 h, which is quite
different from the spectroscopic period $P_{sp}$ = 2.0 h measured by Szkody et al.~(2003). This
appears to be a further case of a measured $P_{sp}$ not agreeing with $P_{orb}$: the previous examples
are (a) GW Lib, in which the photometric period $P_{phot}$ = 2.08 h (Woudt \& Warner 2002b) bears
no obvious relation to the measured $P_{sp}$ = 1.28 h (Thorstensen et al.~2002), (b) FS Aur
in which $P_{sp}$ = 1.428 h and $P_{phot}$ = 3.4247 h (Tovmassian et al.~2003), and (c) HS 2331+3905 where 
$P_{sp}$ is not stable but is always near to 3.5 h and $P_{phot}$ = 1.351 (Araujo-Betancor et al.~2004). 
In the last case there is no doubt, from the presence of eclipse, that $P_{phot} = P_{orb}$.

In Aqr1 we deduce the presence of a magnetic primary rotating with a period of 6.7 min, which eliminates
any interpretation of $P_{sp}$ as arising in a slowly rotating primary.

\subsection{FIRST\,J102347.8+003841 in Sextans}

This star was discovered as a radio source with the use of the Very Large 
Array and was identified in the optical as a 17th magnitude star (Bond et al.~2002). 
It was later independently discovered in the Sloan Digital Sky 
Survey, where it is given the designation SDSS\,1023 (Szkody et al.~2003). 
CVs are not strong radio sources, and the few detections are of systems 
many magnitudes brighter. The spectrum has strong emission lines of He\,II 
4686 {\AA} and of He\,I, as well as the usual Balmer lines, which Bond et al.~suggest 
could be the signature of a magnetic primary, which might go some 
way to explaining the radio flare. Spectra and radial velocities obtained by 
Szkody et al.~(2003) over only 1.7 h suggest a $P_{orb} \sim 3$ h and variations that 
they consider to be more typical of an intermediate polar than a polar.
 
\begin{figure}
\centerline{\hbox{\psfig{figure=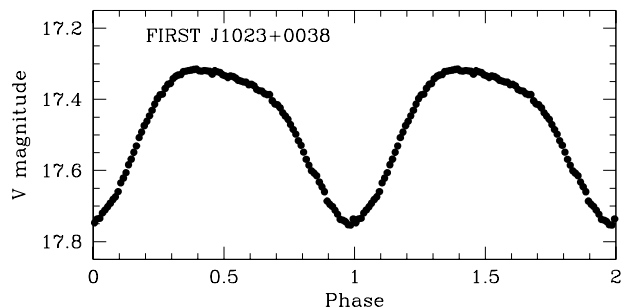,width=8.3cm}}}
  \caption{The average light curve of FIRST\,J1023+0038 (plotted twice), folded on the ephemeris
given in Eq.~\ref{ephfirst}.}
 \label{meanfirst}
\end{figure}

    A CV discovered by such a unique method invites further study. Our 
photometric runs are listed in Table 1, the individual light curves are shown 
in Fig.~\ref{lcfirst} and the binned mean light curve, phased according to the 
ephemeris derived below, is given in Fig.~\ref{meanfirst}.

    FIRST\,J1023+0038 has a strong repetitive modulation with a range of 
0.45 mag about a mean of V $\sim$ 17.5 and a period of 4.7548 h. Low amplitude 
flickering is evident, showing the presence of mass transfer. Fourier 
transforms of our light curves show only the fundamental and first harmonic 
of the principal modulation -- there is no evidence for any other periods in 
the system. The absence of a subharmonic shows that the observed period is 
the actual orbital period. The ephemeris for the times of minima is

\begin{equation}
  {\rm HJD_{min}} = 245\,2668.428 +  0{\fd}198115 (\pm 2) \, {\rm E}.
 \label{ephfirst}
\end{equation}

    The profile of the brightness modulation resembles that of a reflection 
effect, rather than an eclipse. The implication is that there is a hot white 
dwarf in the system, possibly indicative of a nova eruption, overlooked some 
time in the not too distant past (e.g. a few decades ago). A list of reflection 
effects in known recent novae is given in Woudt \& Warner (2003b) in 
connection with the CV RX\,J1039.7--0507. FIRST\,J1023+0038 is only 10 
degrees away in the same constellation, with Galactic coordinates $\ell = 243^\circ$
and $b = +46^\circ$ and is therefore far from the direction of the Galactic bulge 
where most searches for novae tend to concentrate.

   The optimal condition for a large reflection effect is reached for short 
orbital period systems, where the separation of the components is small, 
and/or in absence of screening of the secondary by an optically thick 
accretion disc -- which can only happen if there is a strongly magnetic 
primary (Prialnik 1986; Kovetz, Prialnik \& Shara 1988). At $P_{orb}$ = 4.75 h 
FIRST\,J1023+0038 does not have the required small $P_{orb}$ and therefore is 
probably strongly magnetic, i.e. it is likely to be a polar. If so, it could have 
been desynchronised by the putative nova eruption -- detection of modulated 
polarisation at a period slightly different from that of $P_{orb}$ would confirm our 
suggestion that it is a nova remnant. The pole switching in a desynchronised 
polar (e.g. Mason et al.~1998) and/or reconnection events in the magnetic 
field lines connecting the primary and the secondary could produce just the 
conditions to explain the unusual radio flare in FIRST\,J1023+0038; but in 
any case, the independent deduction of the presence of a strong magnetic 
field is compatible both with the spectrum and the existence of radio 
emission.

\subsection{H$\alpha$\,0242--2802 in Fornax}

The star known currently as H$\alpha$\,0242--2802, which is in the constellation 
Fornax and will be abbreviated here to H$\alpha$\,0242, was discovered in an H$\alpha$ 
emission line survey (Clowes et al.~2002) and found to have a spectrum 
resembling WZ Sge, with double peaked emission lines and the underlying 
white dwarf absorption spectrum, by Howell et al.~(2002), who measured its 
brightness at B = 19.0. It is clearly a CV with a very low rate of mass 
transfer.

\begin{figure}
\centerline{\hbox{\psfig{figure=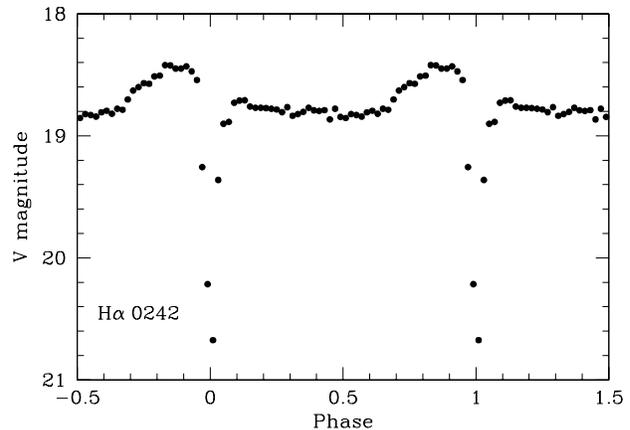,width=8.3cm}}}
  \caption{The average light curve of H$\alpha$\,0242 (plotted twice), folded on the ephemeris
given in Eq.~\ref{ephha0242}.}
 \label{lcha0242}
\end{figure}

   Our photometric runs on H$\alpha$\,0242 are listed in Table 1. H$\alpha$\,0242 is a deeply 
eclipsing system with a period of 1.790 h, showing an orbital hump lasting 
about half the orbital period, interrupted by an eclipse, very much in the 
classical style of Z Cha and OY Car. Our photometric period agrees very well with 
the recently obtained spectroscopic period of $108 \pm 5$ min of H$\alpha$\,0242 (Mason \& Howell 2004).
We acquired observations of four
eclipses on two consecutive nights. The mean light curve is shown in Fig.~\ref{lcha0242} 
and the ephemeris of mid-eclipse is 

\begin{equation}
  {\rm HJD_{min}} = 245\,2577.4376  + 0{\fd}07460  (\pm 1) \, {\rm E}.
\label{ephha0242}
\end{equation}

   H$\alpha$\,0242 is a candidate for searching for non-radial pulsations in its white 
dwarf primary. The spectrum and the eclipse profile indicate that the 
primary contributes at least half of the light in the system; such objects are 
commonly found to have ZZ Cet primaries (Warner \& Woudt 2003).

\subsection{GI Monoceros}

     Nova Monocerotis was discovered at 8.5 mag by M. Wolf in 
February 1918 some weeks after maximum, which probably was 
V $\sim$ 5.2 at the beginning of the year. It was a fast nova, later 
designated as GI Mon. Modern spectra show weak H$\alpha$ emission on 
a very blue continuum (Liu \& Hu 2000), typical of a high $\dot{M}$ 
optically thick accretion disc.

     Prior to eruption a single photographic detection shows it to 
have been at magnitude 15.1, a brightness to which it returned by 
1975  (Robinson 1975). However, there has been a decline since 
then: Szkody (1994) measured V = 16.2 in 1989, which agrees 
with our current estimate (Table 1). This is a greater range than 
could be attributed to the slow ($\sim$ 60 d) oscillations in brightness 
with a range of 0.6 mag, typical of high $\dot{M}$ discs, found to be 
present in GI Mon (Honeycutt 2001).

   Our observations are listed in Table 1 and consist of a two runs 
obtained in 2000 (of which only S6155 is long enough to be 
useful) and three more three years later. The latter were obtained 
with the SAAO CCD photometer mounted on the 1-m reflector, 
which we had access to at the same time that we were using the 
UCT CCD photometer on the 1.9-m telescope. The longer read-out 
time for the SAAO CCD resulted in lower time resolution than 
what we were able to achieve with the UCT CCD.

\begin{figure}
\centerline{\hbox{\psfig{figure=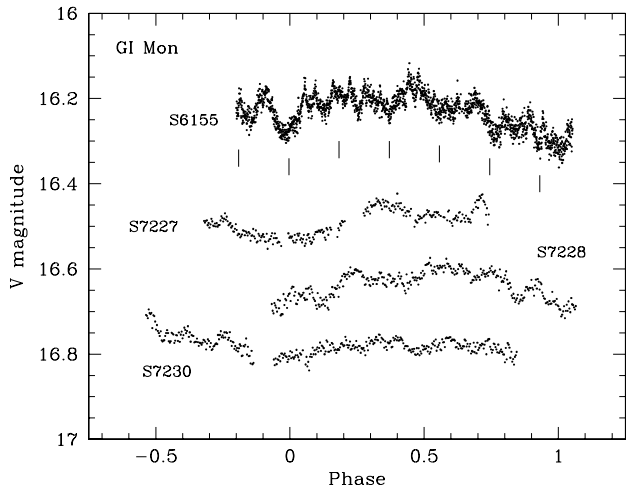,width=8.3cm}}}
  \caption{The light curves of GI Mon, phased according to the ephemeris
given in Eq.~\ref{ephgimon}. Run S7227 is displayed at the correct brightness;
runs S6155, S7228 and S7230 are displaced vertically by -0.1, 0.2 and 0.35 mag, respectively.}
 \label{lcgimon}
\end{figure}

\begin{figure}
\centerline{\hbox{\psfig{figure=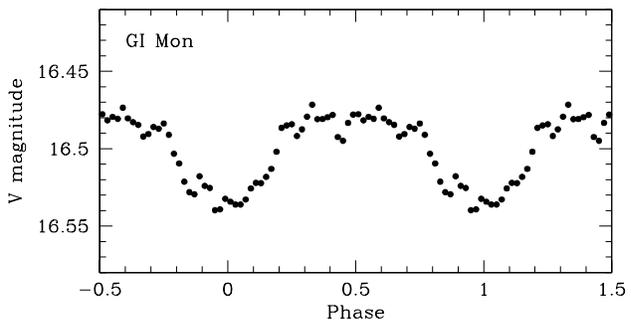,width=8.3cm}}}
  \caption{The average light curve of GI Mon (plotted twice), folded on the ephemeris
given in Eq.~\ref{ephgimon}.}
 \label{lcgimonav}
\end{figure}

   The light curves are shown in Fig.~\ref{lcgimon}, with the 2003/4 group 
aligned according to the period given below. The mean light curve 
for the latter is shown in Fig.~\ref{lcgimonav}. GI Mon shows a brightness 
variation with a range of $\sim$ 0.10 mag and a period of 4.325 h, which 
is indicative of orbital or superhump modulation in a system of moderate 
inclination. The ephemeris for minimum brightness of these 
observations is: 

\begin{equation}
  {\rm HJD_{min}} = 245\,3004.4634  + 0{\fd}1802  (\pm 3) \, {\rm E}.
\label{ephgimon}
\end{equation}

\begin{figure}
\centerline{\hbox{\psfig{figure=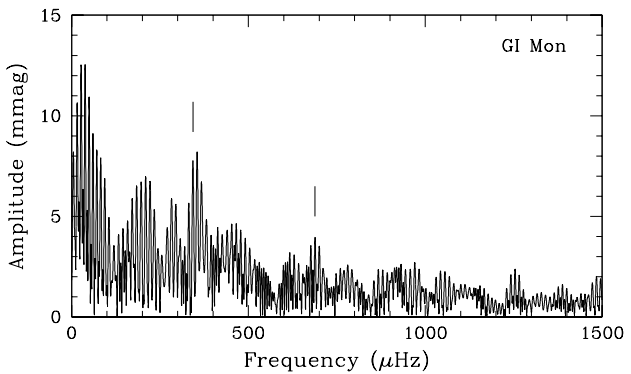,width=8.3cm}}}
  \caption{The Fourier transform of GI Mon, prewhitened at the orbital frequency and its first harmonic.
The peak at 48.6 min and its first harmonic are marked by vertical bars.}
 \label{ftgimon}
\end{figure}

The FT for the 2003/4 light curves, after prewhitening with the 4.325 h modulation
and its first harmonic, shows a peak at 48.6 min (Fig.~\ref{ftgimon}). Within the errors
of the least squares fit, a peak at its first harmonic (at 24.2 min) is identified in the FT;
both the fundamental and first harmonic of this modulation are marked in Fig.~\ref{ftgimon}.
The mean range of the 48.6-min modulation, as seen in the FT, is 16 mmag but individual cycles have much
large amplitudes. 

In the FT of run S6155 (taken three years earlier) 
excess power is seen around 52 min; given the relative short data length ($\sim$ 7 cycles of the
putative 48.6 min modulation), this peak is not well resolved in this individual run. In Fig.~\ref{lcgimon}
we have marked the consecutive minima of the 48.6 min modulation in run S6155 by vertical bars.
There are no shorter period coherent modulations in the light 
curves.

   Our conclusion is that GI Mon is a candidate intermediate polar with an 
orbital period of 4.33 h and a spin-related period of 48.6 min. Our 
observations are not sufficient to detect possible orbital side bands 
to the 48.6 min signal, so we do not know whether it is the white 
dwarf rotation period or a reprocessed signal.

\begin{figure*}
\centerline{\hbox{\psfig{figure=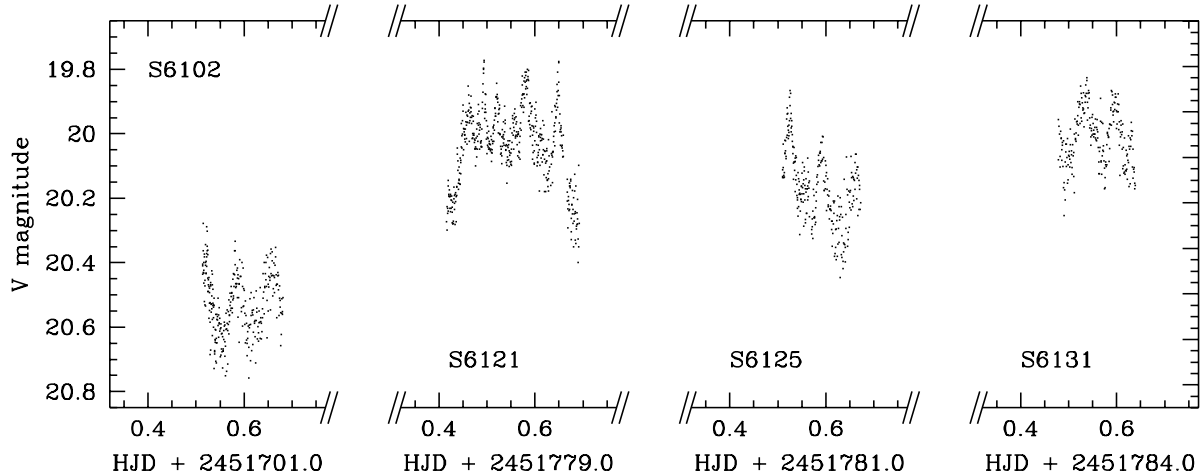,width=16.0cm}}}
  \caption{The light curves of AO Oct, obtained in 2000 June and August.}
 \label{lcaooct}
\end{figure*}

\subsection{AO Octantis}

AO Oct is a dwarf nova of the SU UMa type, showing superoutbursts 
approximately once per year. At maximum of outburst it reaches V $\sim$ 15.3\footnote{The 
General Catalogue of Variable Stars (Kholopov et al.~1985) states that 
maximum is at V = 13.5, and this has been perpetuated in the literature. 
However, we believe this to be a transcriptional error as the original 
discovery paper by von Gessner \& Meinunger (1974) reports a maximum of 
15.3.}; at minimum it can be as faint as V = 21 (Vogt \& Bateson 1982; 
Howell et al.~1991). Superhumps with a period of 96.70 min are observed 
during superoutburst (Patterson et al.~2003). Early photometric observations 
of AO Oct at quiescence failed to show any orbital modulation (Howell et al.~1991), 
but those data were very noisy and at the limit of the instrument 
available (V = 20.9, observed with a 1-m telescope and a V filter). More 
recent photometry (September 2000) has been successful (Patterson et al.~2003).

    Our photometric observations are listed in Table 1. We detected orbital 
modulation in our first observation in June 2000 and later extended coverage 
with three nights in August 2000, when it happened that AO Oct was about 
0.5 mag brighter than in June. The light curves are shown in Fig.~\ref{lcaooct}; an 
orbital modulation with a peak-to-peak range of 0.2 mag is clearly seen, but 
there was also a great deal of flickering during run S6121.
   The FT of the August light curves has a peak at 94.10 min, which we 
increased in accuracy by performing a non-linear least squares fit to find a 
period of 94.097 ($\pm$ 0.022) min. This is in reasonable agreement with the 
value 94.43 ($\pm$ 0.19) min obtained from photometry at quiescence by Patterson 
et al.~(2003). The ephemeris for maximum light is given by

\begin{equation}
  {\rm HJD_{max}} = 245\,1781.5261  + 0{\fd}065345  (\pm 15) \, {\rm E}.
\label{ephaooct}
\end{equation}

\begin{figure}
\centerline{\hbox{\psfig{figure=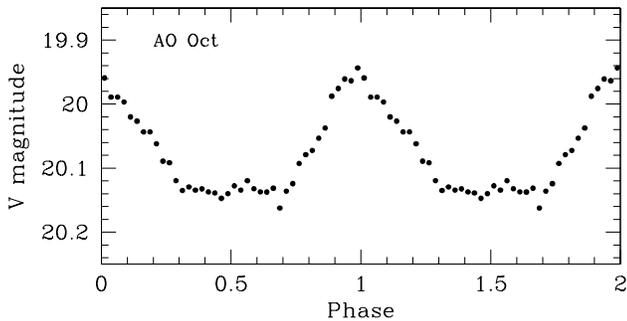,width=8.3cm}}}
  \caption{The mean light curve of AO Oct in August 2000 (plotted twice), folded on the ephemeris given in Eq.~\ref{ephaooct}.}
 \label{meanaooct}
\end{figure}

   Fig.~\ref{meanaooct} shows the mean light curve obtained by folding our light curves 
at the orbital period. The superhump period is 2.8\% longer than the orbital 
period, giving a beat period between the two of 2.4 d, which is typical of SU 
UMa stars of this orbital period (Warner 1995).

\subsection{V972 Ophiuchi}

V972 Oph was Nova Ophiuchi 1957, discovered by Haro at magnitude 9.8 
(Haro 1957), but reaching 8.0 at maximum. It is classified as a slow nova.
The spectrum obtained by Ringwald et al.~(1996) shows a red continuum 
with no distinct emission lines; they measured its brightness as V = 16.7. 
However, Zwitter \& Munari (1996) measured B = 17.5 and although they 
also saw a red continuum, they found double-peaked He\,II emission, strong 
H$\alpha$ emission, and weak H$\beta$ emission superimposed on wide absorption, 
typical of a high $\dot{M}$ disc. The absence of He\,I lines led them to conclude 
that the white dwarf in V972 Oph is still extremely hot. The continuum slope is 
due to reddening in this low galactic latitude ($b = +2^\circ$) nova.

    The spectra confirm that the nova remnant has been correctly identified 
(Duerbeck 1987) but our first photometric observation, made in April 2002, 
showed no sensible short time scale variations. Revisiting in August 2003 
revealed the expected activity. It is possible that the changes in level of 
activity are correlated with the strength of the emission lines, and that the 
spectrum obtained by Ringwald et al.~is characteristic of the quieter 
moments.

\begin{figure}
\centerline{\hbox{\psfig{figure=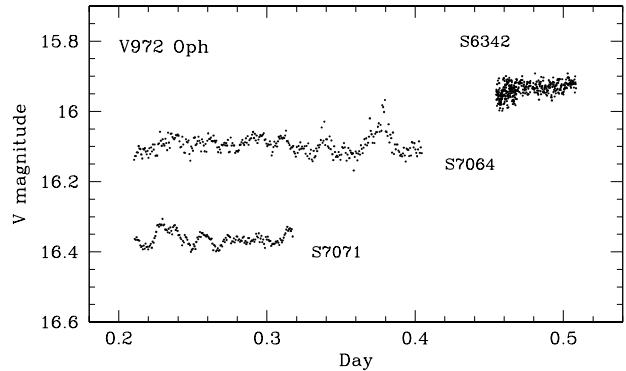,width=8.3cm}}}
  \caption{The light curves of V972 Oph. The light curve of run S7071 is displaced vertically downwards by 0.25 mag for
display purposes.}
 \label{lcv972oph}
\end{figure}

     Our runs are listed in Table 1 and the light curves are shown in Fig.~\ref{lcv972oph}. 
The 2003 runs show considerable activity, but there is no sign of any orbital 
modulation on time scales of hours and the FTs do not show any persistent 
periodicities.

\subsection{SDSS\,J015543.40+002807.2 in Cetus}

SDSS\,0155 is a CV announced in the first Sloan Digital Sky Survey release (Szkody et 
al.~2002), where its strong He\,II emission and its identification as a ROSAT source clearly 
pointed to its being a polar. Comparison of SDSS photometry and earlier photographic 
images showed that SDSS 0155 has high and low states of luminosity, over the range 
14.7 -- 17.6 mag, compatible with a polar classification. Radial velocity measurements of 
the HeII lines obtained when the system was in a high state gave an orbital period of 
88 $\pm$ 2 min.

\begin{figure}
\centerline{\hbox{\psfig{figure=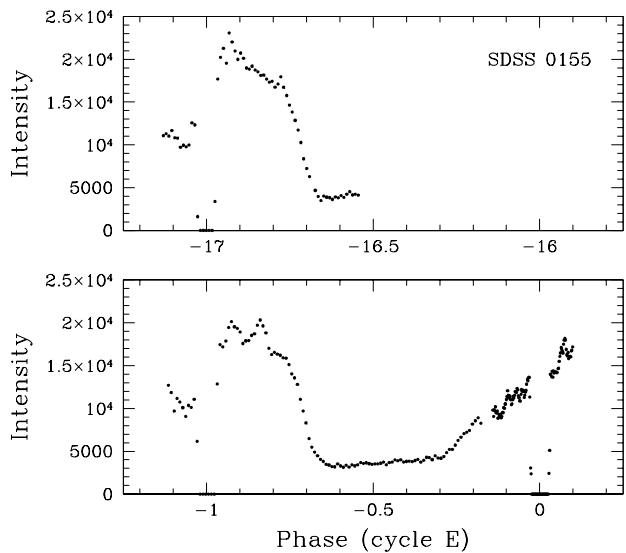,width=8.3cm}}}
  \caption{The light curves of SDSS\,0155 (plotted in intensity), aligned according to the ephemeris
given in Eq.~\ref{ephsdss0155}.}
 \label{lcsdss0155}
\end{figure}

   Our photometric observations are listed in Table 1 and were obtained when SDSS\,0155 
was in a low state. The light curves are given in Fig.~\ref{lcsdss0155} and show the star to be a 
deeply eclipsing system with an eclipse lasting $\sim$ 320 s and a period of 87.13 $\pm$ 0.02 min. 
Because it is too faint to detect when in eclipse we have plotted the light curve on an 
intensity instead of a magnitude scale. We have used zero intensities (rather than noise-determined 
upper limits) for the integrations obtained in eclipse. The ephemeris for the 
middle of eclipse of the primary is

\begin{equation}
     {\rm HJD_{min}} = 245\,2918.42776 + 0.06051\,(1)\, {\rm E}
\label{ephsdss0155}
\end{equation}

   The light curve shows that SDSS\,0155 is probably a single pole accretor, with the 
accretion region hidden behind the primary for roughly half an orbital period.

\subsection{SDSS\,J023322.61+005059.5 in Cetus}

SDSS\,0233 is another CV in the first release of the SDSS (Szkody 
et al.~2002), where its spectrum was found to be dominated by 
narrow Balmer emission and weak He I on a blue continuum. In 
appearance it possibly resembles a low inclination dwarf nova in 
quiescence, but Szkody et al.~found it to be a highly variable X-Ray 
source, which is uncharacteristic of a CV with such a low energy 
optical spectrum.

\begin{figure}
\centerline{\hbox{\psfig{figure=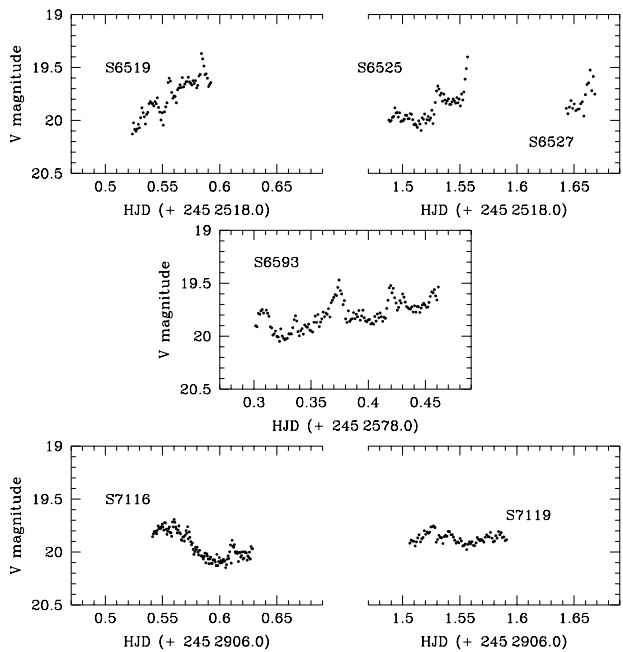,width=8.3cm}}}
  \caption{The light curves of SDSS\,0233.}
 \label{lcsdss0233}
\end{figure}

   Our observations are listed in Table 1 and the light curves are 
displayed in Fig.~\ref{lcsdss0233}. The star shows slow flaring activity of 
considerable amplitude, and faster flickering, but despite our 
having devoted a moderate amount of photometric time to it there 
are no hopeful indications of any persistent periodicity, so we have 
relegated it.

\subsection{SDSS\,J124058.03-015919.2 in Virgo}

SDSS\,1240 was discovered recently as a candidate AM CVn star following a systematic search for AM CVn systems 
in the first data release of the Sloan Digital Sky Survey (Roelofs et al.~2004). SDSS\,1240
was previously identified as a DB white dwarf following the 2dF quasar redshift survey
(Croom et al.~2001) -- the 2dF spectrum shows broad absorption lines in the blue part of the spectrum.
Roelofs et al.~(2004) have obtained spectra with the 6.5-m Magellan telescope and clear double-peaked He\,I emission
lines are present, as well as He\,II (4686 {\AA}) and N\,III (4634 and 4640 {\AA}) emission lines.

\begin{figure}
\centerline{\hbox{\psfig{figure=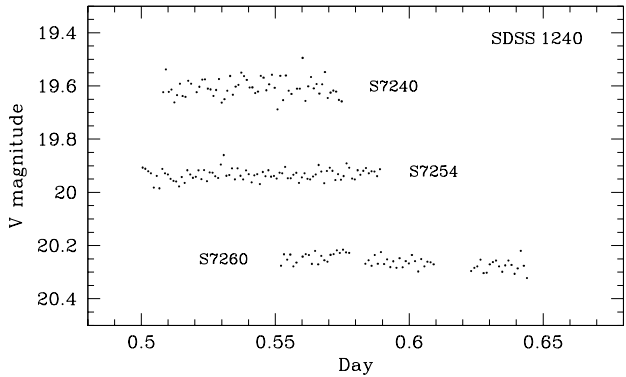,width=8.3cm}}}
  \caption{The light curves of SDSS\,1240. Run S7240 is displayed at the correct brightness; runs S7254 and S7260 are displaced
vertically downwards by 0.3 and 0.6 mag, respectively.}
 \label{lcsdss1240}
\end{figure}

Our photometry is listed in Table 1 and the longer light curves (runs S7240, S7254 and S7260) are shown in
Fig.~\ref{lcsdss1240}.  Surprisingly, no variations are seen in the light curves and combined FT. 
There is little or no flickering, and no orbital modulation is detectable. The orbital period will have
to be determined from spectroscopy.

\subsection{SDSS\,J155644.24-000950.2 in Serpens}

SDSS\,1556 emerged as a V $\sim$ 18.1 mag star in the first release of Sloan 
Digital Sky Survey CVs (Szkody et al.~2002) where the spectrum is seen to 
have Balmer emission lines, with no He\,II emission, and the H$\beta$ emission is 
centred in the absorption trough of the underlying white dwarf spectrum. 
These are the characteristics of a dwarf nova with low $\dot{M}$, probably of 
short orbital period.

\begin{figure}
\centerline{\hbox{\psfig{figure=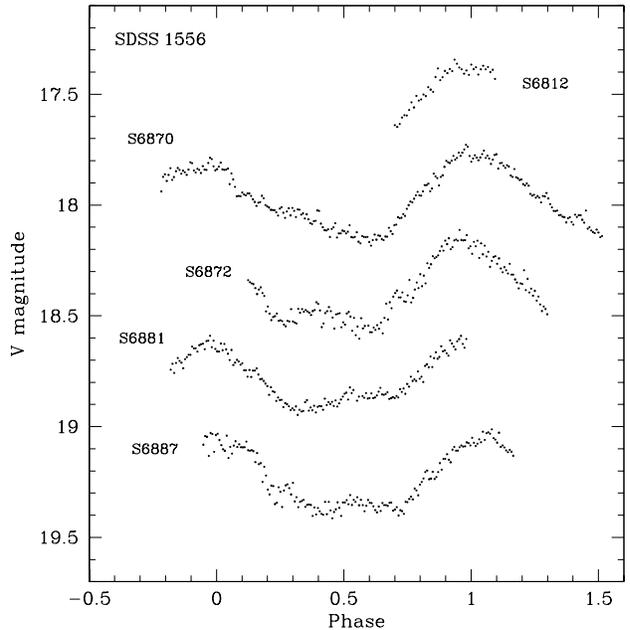,width=8.3cm}}}
  \caption{The light curves of SDSS\,1556, folded on the ephemeris given in Eq.~\ref{ephsdss1556}. 
Run S6870 is displayed at the correct brightness. Runs S6872, S6881 and S6887 are
vertically displaced by 0.4, 0.8 and 1.2 mag, respectively. Run S6812 is displaced upwards by -0.4 mag and arbitrarily shifted
in phase to match the time of maximum.}
 \label{lcsdss1556}
\end{figure}

\begin{figure}
\centerline{\hbox{\psfig{figure=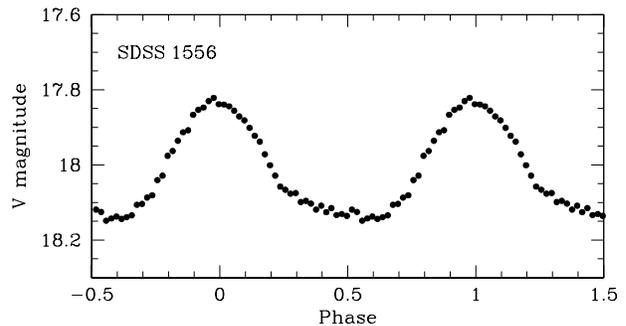,width=8.3cm}}}
  \caption{The average light curve of SDSS\,1556, folded on the ephemeris
given in Eq.~\ref{ephsdss1556}.}
 \label{meansdss1556}
\end{figure}

\begin{figure}
\centerline{\hbox{\psfig{figure=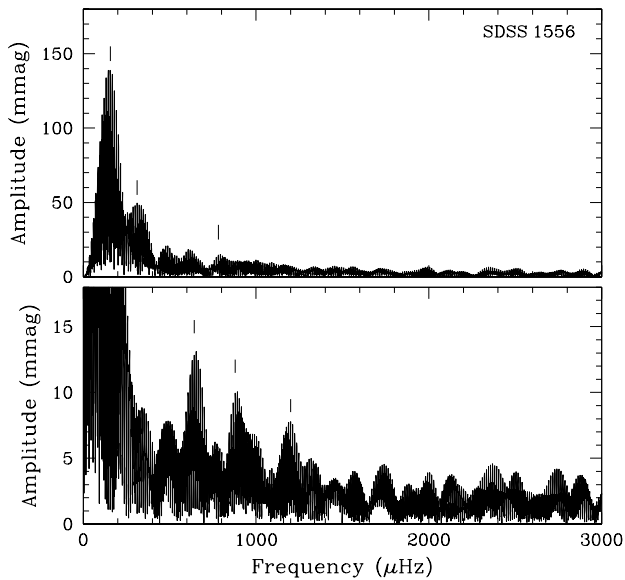,width=8.3cm}}}
  \caption{The Fourier transform of the 2003 March observations of SDSS\,1556. The upper panel shows the total FT with the 
orbital frequency marked (and the first and fourth harmonic). The lower panel shows the total FT after prewhitening at the orbital
frequency and the two harmonics.}
 \label{ftsdss1556}
\end{figure}

   Our observations are listed in Table 1 and the light curves are shown in 
Fig.~\ref{lcsdss1556}. There is a clear non-sinusoidal modulation with a period of
$\sim$ 1.78 h, a peak-to-peak range of 0.4 mag. and small variations in profile from 
cycle to cycle. The mean light curve, summed modulo the measured period, 
is shown in Fig.~\ref{meansdss1556}. The FT of the light curves, seen in Fig.~\ref{ftsdss1556}, has an
aliasing ambiguity at the fundamental frequency, but the presence of the first (and fourth)
harmonic gives further estimates of the period, and there is 
only one alias at the fundamental that furnishes fits (to within $\sim$ 1 $\mu$Hz) to 
the aliases in the harmonic patterns. This gives a period of 1.778 h (106.7 
min). The ephemeris for maximum light is:

\begin{equation}
  {\rm HJD_{max}} = 245\,2725.5160 + 0{\fd}07408  (\pm 1) \, {\rm E}.
\label{ephsdss1556}
\end{equation}

   It is probable that this period is $P_{orb}$ for SDSS\,1556, but the profile of the 
mean light curve (Fig.~\ref{meansdss1556}) is similar to that of the long outburst interval 
dwarf nova GW Lib (Woudt \& Warner 2002b) which we have discussed in Sect.~2.2,
so we advise caution in interpretation of the photometric period.

The FT prewhitened at the fundamental and first and fourth harmonic shows low amplitude but significant structure,
with peaks near 1558 s, 1137 s, and 834 s (the lower panel of Fig.~\ref{ftsdss1556}); the low frequency peak
off the scale in the prewhitened FT is due to the data length of some observing runs. As SDSS\,1556 is a low $\dot{M}$
system it could be expected to show non-radial oscillations of the ZZ Cet type, as detected in a number of other
low $\dot{M}$ CVs (Warner \& Woudt 2003). Our observed periods are somewhat beyond the observed long end of non-accreting
ZZ Cet stars, but are theoretically possible (see discussion by Mukadam et al.~2002). We need to obtain further observations
before drawing a conclusion on the nature of these periodicities.

\subsection{SDSS\,J205017.84-053626.8 in Aquarius}

SDSS 2050 is another product of the second release of CVs in the Sloan Digital Sky Survey (Szkody et al.~2003). The
spectrum has very prominent He\,II 4686 {\AA}, almost as strong as neighbouring H$\beta$, which is a frequent characteristic of
magnetic CVs. Szkody et al.~(2003) find a tentative $P_{orb}$ of $\sim$ 2 h from only 1.8 h of spectral coverage. Of
considerable significance is their observation of a low state in July 2002, compared with a month later when it was
bright enough to obtain the spectra; again this is characteristic of magnetic CVs -- in particular, polars.

\begin{figure}
\centerline{\hbox{\psfig{figure=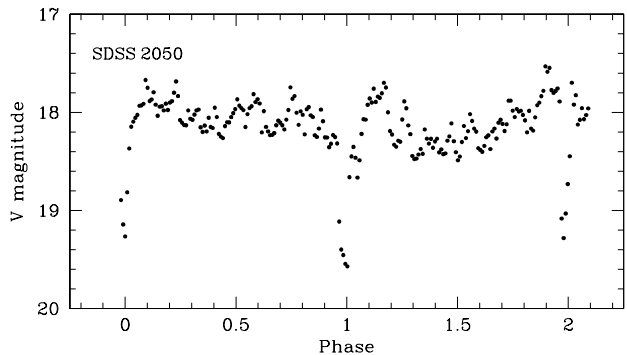,width=8.3cm}}}
  \caption{The light curve of SDSS\,2050, phased according to the ephemeris given in Eq.~\ref{ephsdss2050}.}
 \label{lcsdss2050}
\end{figure}

Our observations of SDSS 2050 are listed in Table 1 and the longest light curve (S7062) is shown in Fig.~\ref{lcsdss2050}.
The other runs were obtained around the times of eclipses in order to improve the period determination. The light
curve is clearly that of a high inclination polar, with eclipses $\sim$ 1.5 mag deep of $\sim$ 260 s duration
recurring with a period $P_{orb}$ = 1.5702 h. The ephemeris for mid eclipse is given by

\begin{equation}
  {\rm HJD_{min}} = 245\,2880.32762 +  0{\fd}0654246 (\pm 5) \, {\rm E}.
 \label{ephsdss2050}
\end{equation}

\subsection{FH Serpentis}

\begin{figure}
\centerline{\hbox{\psfig{figure=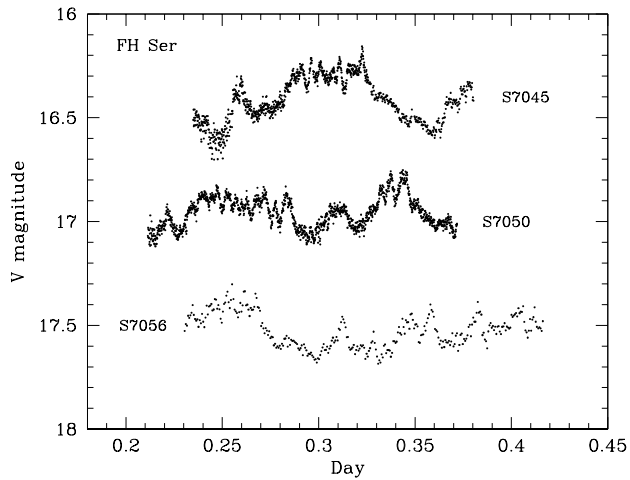,width=8.3cm}}}
  \caption{The light curves of FH Ser. Runs S7045 and S7056 are displaced vertically by -0.7 mag and +0.5 mag, respectively. Run 
S7050 is displayed at the corrected brightness.}
 \label{lcfhser}
\end{figure}

FH Ser was Nova Serpentis 1970, discovered by Hirose \& Honda (1970) 
and, during eruption, was one the best-studied novae of the twentieth 
century, showing a great increase in infrared luminosity at the time that the 
optical brightness diminished through the formation of dust in the ejecta. It 
reached V = 4.4 at maximum and has declined to post-outburst brightness of 
V $\sim$ 17.6. Its modern classification is a slow nova of the Fe\,II class (Williams 
1992). It has an observed ejecta shell (Gill \& O'Brien 2000). Despite its 
importance in the history of understanding the evolution of nova light 
curves, it does not have a known orbital period, and no high speed 
photometry for it has been published. The spectrum in 1991 was still 
dominated by emission lines from the ejecta (Ringwald et al.~1996). 
We observed FH Ser in the hope of detecting an orbital brightness 
modulation. Our observations are listed in Table 1 and the light curves are 
displayed in Fig.~\ref{lcfhser}.  

    FH Ser shows rapid flickering superimposed on slow excursions of $\sim$ 0.5 
mag on time scales $\sim$ 2.5 h. The FTs do not show evidence for any coherent 
periodicity on time scales up to a few hours. The large amplitude of 
flickering will mask orbital modulation; to obtain the orbital period by a 
photometric method will probably require a multi-site collaborative 
campaign.

\section{Summary and discussion}

    Our observations are of a mixed group of CVs -- four known old novae 
(GI Mon, V356 Aql, V972 Oph and FH Ser), in the latter three we found no orbital 
modulation, and a possibly overlooked nova (FIRST\,J1023). We have found 
two new polars with deep eclipses (SDSS\,0155 and SDSS\,2050) and a new intermediate polar 
(Aqr1). We have determined orbital periods for two dwarf novae in 
quiescence (AO Oct and H$\alpha$\,2042) and observed periodic 
modulations in one other dwarf novae (SDSS\,1556) that 
might be orbital but could also be a possible further manifestation of the GW Lib 
phenomenon of a non-orbital modulation of unknown origin. We find no periodic
photometric signal in SDSS\,0233, nor in the new AM CVn star SDSS\,1240.

    Among our results is one CV, Aqr1, for which the spectroscopic
period differs considerably from the photometric period. We summarise this in Table 2
and draw attention to the clustering of period ratios, which is perhaps indicative of
some resonance effect in operation.

\begin{table}
 \centering
  \caption{Stars showing the GW Lib phenomenon.}
  \begin{tabular}{@{}cccc@{}}
Star                    & $P_{phot}$ (h)    & $P_{sp}$ (h)  & $P_{phot}/P{sp}$ \\[10pt]
GW Lib       &  2.09    & 1.280   & 1.63 \\
Aqr1         &  3.226   & 2.0     & 1.6  \\
FS Aur       &  3.425   & 1.428   & 2.40 \\
HS\,2331     &  1.351   & 3.5     & 1/2.6 \\
\end{tabular}
\label{tab2}
\end{table}

\section*{Acknowledgments}
We thank Dr. D.~O'Donoghue for the use of his EAGLE program for 
Fourier analysis of the light curves. 
PAW is supported by funds made available from the National Research
Foundation and by strategic funds made available to BW from the
University of Cape Town. BW's research is supported by the University.
MLP is supported by a bursaries from the Department of Labour and the National Research Foundation.

\end{document}